\documentclass{mem}
\usepackage{natbib}\usepackage{txfonts}\usepackage{balance}
\usepackage{graphicx}
\usepackage[a4paper,breaklinks,dvipdfm]{hyperref}
\idline{75}{282}
\begin{document}

\title{Shedding Light on Lithium Evolution}

   \subtitle{The Globular Cluster Perspective}

\author{
Andreas J.\ Korn
          }


\institute{
Department of Physics and Astronomy, Uppsala University, Box 516, 75120 Uppsala, Sweden\\
\email{andreas.korn@physics.uu.se}
}

\authorrunning{A.J.\ Korn}

\titlerunning{Shedding Light on Lithium Evolution}

\abstract{I shall review what has been learnt during 20 years of lithium observations in
stars belonging to metal-poor globular clusters. The focus will be on little evolved main-
sequence, turnoff-point (TOP) and subgiant-branch (SGB) stars expected to display Spite-plateau lithium abundances like those found in the majority of field stars of similar metallicities. But is the Spite plateau of globular clusters the same as those of field stars? What effect does, e.g., cluster-internal pollution have on lithium abundances in the now dominant second generation of stars? It will be shown that it is primarily our incomplete knowledge of the temperature
scale of Population II stars which currently limits the diagnostic power of globular clusters as regards the stellar-surface evolution of lithium.
}
\maketitle{}

\section{Introduction}

Owing to the publish-or-perish policy, all astro\-nomers are good at presenting scientiﬁc re-sults in papers. However, I would argue we are less good at reporting the process of scientiﬁc inference: how do we formulate scientiﬁc questions and how do we go about addressing them? The answers to these questions consist of more than the capabilities of our theories, our simulations, our telescopes and instruments.\footnote{If our ``creations'' were self-aware, they would ask ``Why did you create me?''. Is our only answer ``Because it was possible.''? Cf. antiquity's Prometheus and Mary Shelley's Frankenstein.}
There is human curiosity and scientiﬁc instinct, there are preconceptions and prejudice. I feel that review talks and proceedings can serve a purpose in this context. Here, one can talk about past developments and their influence on the present state of affairs and future directions. As philosopher George Santayana (1863-1952) put it: ``Those who cannot remember the past are condemned to repeat it.'' So let's take a look at 20 years of lithium studies in Galactic Globular Clusters.

\section{Heroic early efforts}
To my knowledge, the first spectra of lithium in a Globular Cluster (GC) star that has not undergone the first dredge-up were taken in June 1992. A 200\AA-wide region covering the 6707\,\AA\ Li I resonance doublet was observed with EMMI on NTT at resolving power $R$ $\approx$ 28,000
and signal-to-noise ratio (S/N) of around 35 (Molaro \& Pasquini 1994). This first star was a warm subgiant in the metal-poor GC NGC 6397 ([Fe/H]=$-2$). Apart from lithium, only one more spectral line was detected: H$\alpha$. The analysis of both lines resulted in a lithium abundance of log $\varepsilon$(Li) = 2.35 $\pm$ 0.25, said to be somewhat higher than lithium abundances in Spite-plateau field stars, but fully compatible with these to within the error bars. 

Deliyannis, Boesgaard \& King (1995) investigated lithium in three warm subgiants of the very metal-poor GC M 92 ([Fe/H]=$-$2.4) using HIRES on Keck. While the resolving power was high ($R$ $\approx$ 45,000), the S/N was low (16 $\leq$ S/N $\leq$ 24). Admittedly, at $V$ magnitudes around 18 these stars are bloody faint for high-resolution studies, even on a 10m telescope. The spread in lithium abundance discovered, 1.92 $\leq$ log $\varepsilon$(Li) $\leq$ 2.33, was ascribed to “different stellar surface Li depletion histories”, i.e. processes not captured by standard stellar evolution theory. Boesgaard et al.\ (1998) extended this sample to seven stars, three with S/N of 40. The full range in lithium now reached a factor of three (0.5\,dex in log(abundance)).

Later, Bonifacio (2002) reanalysed the same data set and found that the data quality does not allow to claim the existence of a dispersion in lithium (a dispersion as large as 0.18\,dex seemed nonetheless compatible with the data). Ten years on, we still do not know the run of lithium among M92 subgiants and its dispersion. Does it really take the TMT to go to the bottom of this? To me, it rather looks like a matter of priorities, especially after the HIRES upgrade.

\section{More lithium in NGC 6397}
\begin{table*}
\caption{Studies since 1994 dedicated to lithium in NGC 6397 TOP/SGB stars}
\begin{center}
\begin{tabular}{lcr}
\hline
\\
authors & log $\varepsilon$(Li) $\pm$ $1\sigma$ & no.\ of stars\\
\hline
\\
Molaro \& Pasquini 1994 & 2.35 $\pm$ 0.25 & 1\\
Pasquini \& Molaro 1996 & 2.28 $\pm$ 0.10 & 3\\
Th\'{e}venin et al.\ 2001 & 2.21 $\pm$ 0.07 & 7\\
Bonifacio et al.\ 2002 & 2.34 $\pm$ 0.06 & 12\\
Korn et al.\ 2007 & 2.24 $\pm$ 0.05 & 7\\
Lind et al.\ 2009a & 2.27 $\pm$ 0.005 & 346\\
Gonz\'{a}lez Hern\'{a}ndez et al.\ 2009 & 2.37 $\pm$ 0.01 & 84\\
Nordlander et al.\ 2012 & 2.26 $\pm$ 0.05 & 7\\
\hline
\end{tabular}
\end{center}
\end{table*}

As one of the most nearby low-metallicity, low-reddening GCs, NGC 6397 has always been a prime target for ESO telescopes. Pasquini \& Molaro (1996) returned to this cluster, this time observing three warm subgiants (much like Deliyannis, Boesgaard \& King did; seemingly, three stars were considered to be a statistically signiﬁcant sample in those days).

The VLT era has added no less than six comprehensive lithium studies on this cluster, initially conducted with UVES (Th\'{e}venin et al.\ 2001, Bonifacio et al.\ 2002), later with FLAMES-UVES (Korn et al.\ 2006, 2007, Nordlander et al.\ 2012) and FLAMES-GIRAFFE (Lind et al.\ 2009a, Gonz\'{a}lez Hern\'{a}ndez et al.\ 2009), all focussing on stars that have not undergone the first dredge-up. Table 1 lists the resulting mean lithium abundances and the standard deviation of the mean. A clear trend towards smaller error bars is apparent. Furthermore, a 0.1\,dex dichotomy in the mean values bet\-ween Mediterranean and Central/Northern-European studies strikes the eye. This difference is predominantly caused by differences in the adopted effective-temperature scale (departures from LTE also play in). The Mediterraneans seem to like it hot! 

Given $T_{\rm eff}$-scale differences exceeding 100\,K (NB: 100\,K corresponds to 0.07\,dex in log $\varepsilon$(Li)), it is obvious that the standard error of the mean of the recent studies is a marked underestimation of the total error budget for the absolute lithium abundance. For the same reason, it is not easy to draw firm conclusions about whether or not NGC 6397 stars have as much (or rather, as little!) lithium as field stars of the same metallicity. Nonetheless, there is little evidence for systematically higher lithium abundances in GCs. Given the different magnitudes of local subdwarfs and GC subgaints, fully differential spectroscopic analyses are rare. 

Based on HST photometry, Milone et al.\ (2012) were able to trace two main sequences in NGC 6397, like in several other clusters previously investigated. It would be interesting to carefully study the chemistry of these two generations of stars, beyond indirect inferences on helium and the usual anti-correlations. Lithium should be looked at to quantify the effect of intra-cluster pollution on this fragile element. For NGC 6397, the effect is likely very small ($\leq$ 0.05\,dex), but it is more prominent in other clusters (e.g. NGC 6752, Pasquini et al.\ 2005). We will ultimately need to understand these trends in terms of stellar and GC structure and evolution.

\section{The role of atomic diffusion and extra mixing}
Atomic-diffusion theory of inhomogeneous gases coupled to stellar structure and evolution calculations predicts large modifications to the initial surface abundances of stars of spectral types A and B. Among later spectral types, it is in particular old, metal-poor F-type stars that are affected, as high ages and shallow surface convection work in tandem to produce sizable effects (Richard, these proceedings). Even if no observational abundance trends compatible with atomic-diffusion predictions were reported in the literature before 2006 (Korn et al.\ 2006), it was implicit\-ly clear to observers that heavy-element abundance variations larger than $\sim$ 0.3\,dex are unlikely to be present when comparing TOP and red-giant-branch (RGB) stars drawn from the same stellar population. Uninhibited atomic diffusion does not seem to be realized in stellar envelopes.

Theorists met this observational constraint by introducing a parametrized form of extra mixing (sometimes called turbulent mixing hinting at the underlying physics). In essence, the structure of the model star is modified by mixing a certain depth range. The density dependence of this extra mixing was calibrated on the Sun ($\rho^{-3}$), the main free parameter is thus its overall strength relative to atomic diffusion (the mixing is in itself modelled as a diffusive process). Proffit \& Michaud (1991) chose to designate models by the logarithm of the reference temperature $T_0$ where the extra-mixing diffusion constant $D_T$ is chosen to be 400 times the atomic-diffusion constant for helium. In this parameter, the range of models fulfilling the observational constraints of a thin and flat Spite plateau is roughly T5.95 to T6.25 (Richard, Michaud \& Richer 2005). These are the models we have explored in Korn et al.\ (2006, 2007), Lind et al.\ (2008, 2009a) and Nordlander et al.\ (2012).

Extra mixing takes care of two things: it moderates the effects of atomic diffusion for all elements, atoms and ions cannot move freely through layers where convection and/or extra mixing is present. For lithium, it provides a layer below the convective envelope where lithium can be stored, i.e. preventing it from being destroyed in deeper, hotter layers. This gives a potentially powerful observational signature of lithium diffusion, as this stored lithium can be mixed back to the surface just before surface dilution in connection with the first dredge-up takes place in the middle of the subgiant branch. See Figs.\ 3 \& 4 of Nordlander et al., these proceedings. 

One perfidious aspect of models with high extra-mixing efficiency: they produce rather small effects for heavy elements ($\sim$ 0.1\,dex), but produce the largest abundance corrections for lithium (up to 0.4\,dex). Keep this in mind for the next section.   

Our current best estimate of the diffusion-corrected initial lithium abundance of Spite-plateau stars in NGC 6397 is log $\varepsilon$(Li) = 2.57 $\pm$ 0.1 (Nordlander et al.\ 2012). We are not more than 1.2 $\sigma$ away from agreement with the WMAP-calibrated Big-Bang nucleosynthesis (BBN) value, 2.71 $\pm$ 0.06 (Cyburt et al.\ 2010).
 
\section{NGC 6752, M 4, M 30...}
\begin{figure*}[t!]
\hspace*{-0.06\hsize}
\resizebox{1.09\hsize}{!}{\includegraphics[clip=true, angle=90]{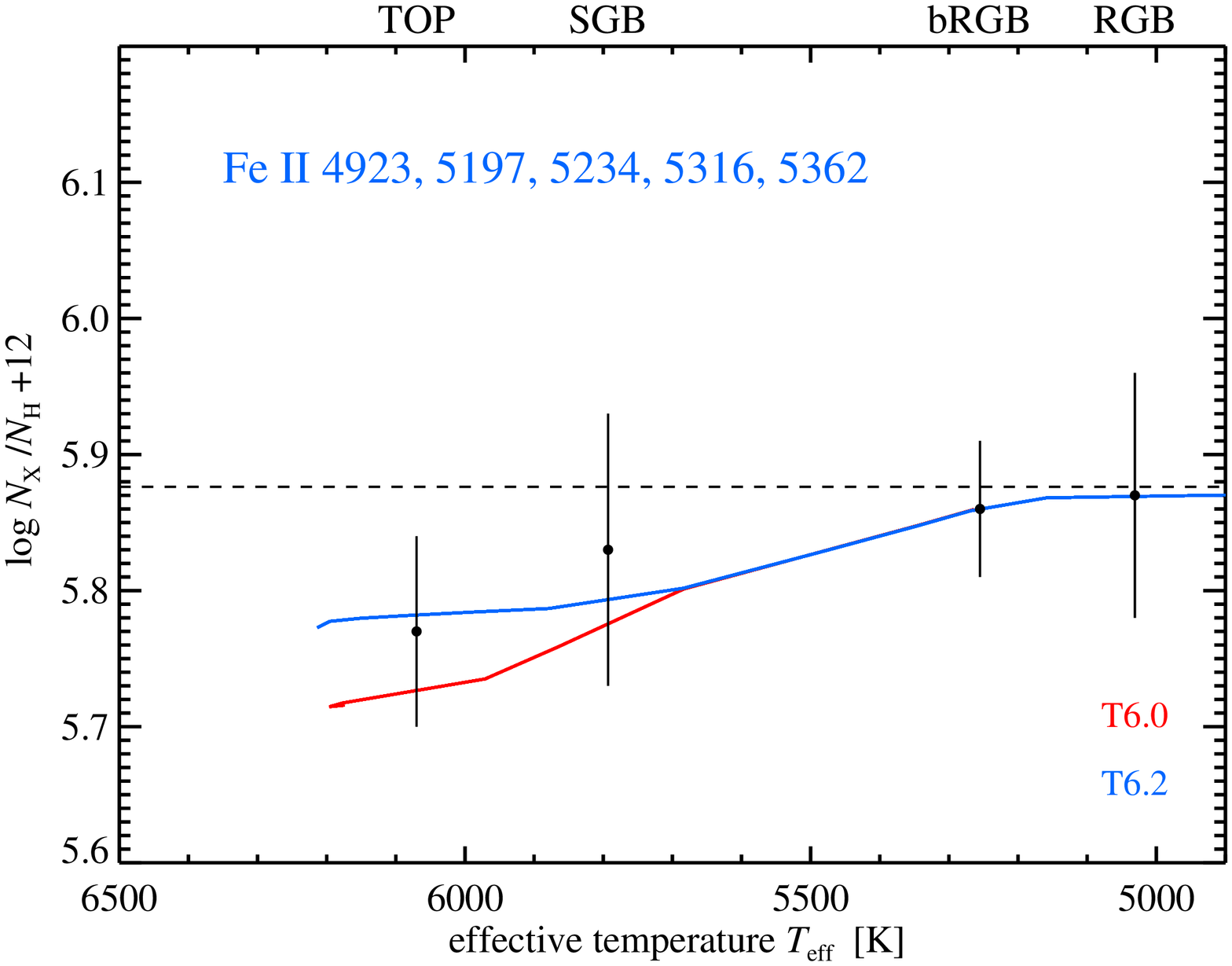}}
\hspace*{-0.06\hsize}
\resizebox{1.09\hsize}{!}{\includegraphics[clip=true, angle=90]{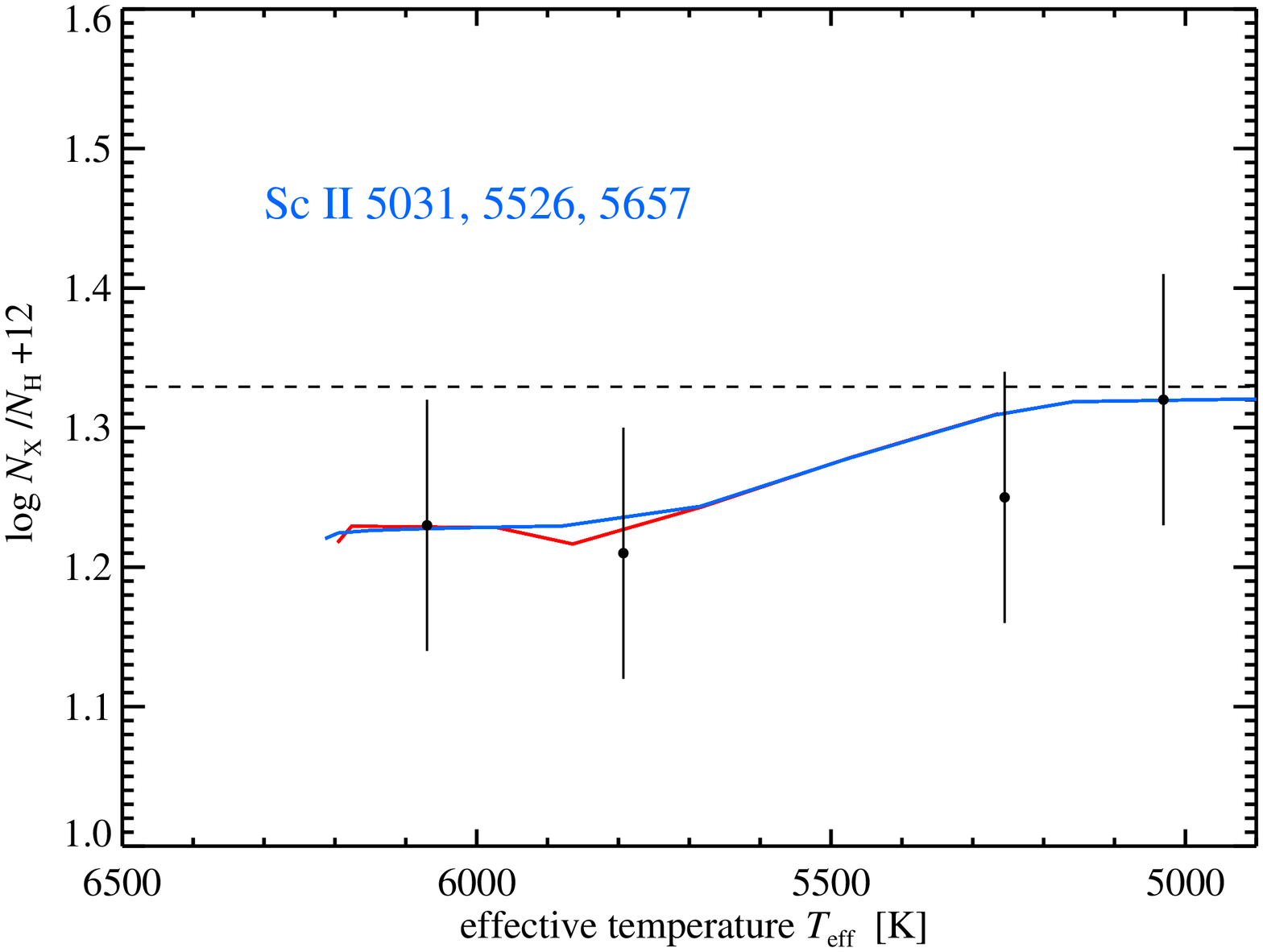}}
\caption{Observed abundance trends for iron and scandium in NGC 6752, based on five and three ionic lines, respectively. Atomic-diffusion model predictions with NGC 6397-like (T6.0, red) and more efficient (T6.2, blue) extra mixing are overplotted (Richard, priv.\ comm.). 
}
\end{figure*}
\begin{figure*}[t!]
\hspace*{-0.06\hsize}
\resizebox{1.09\hsize}{!}{\includegraphics[clip=true, angle=90]{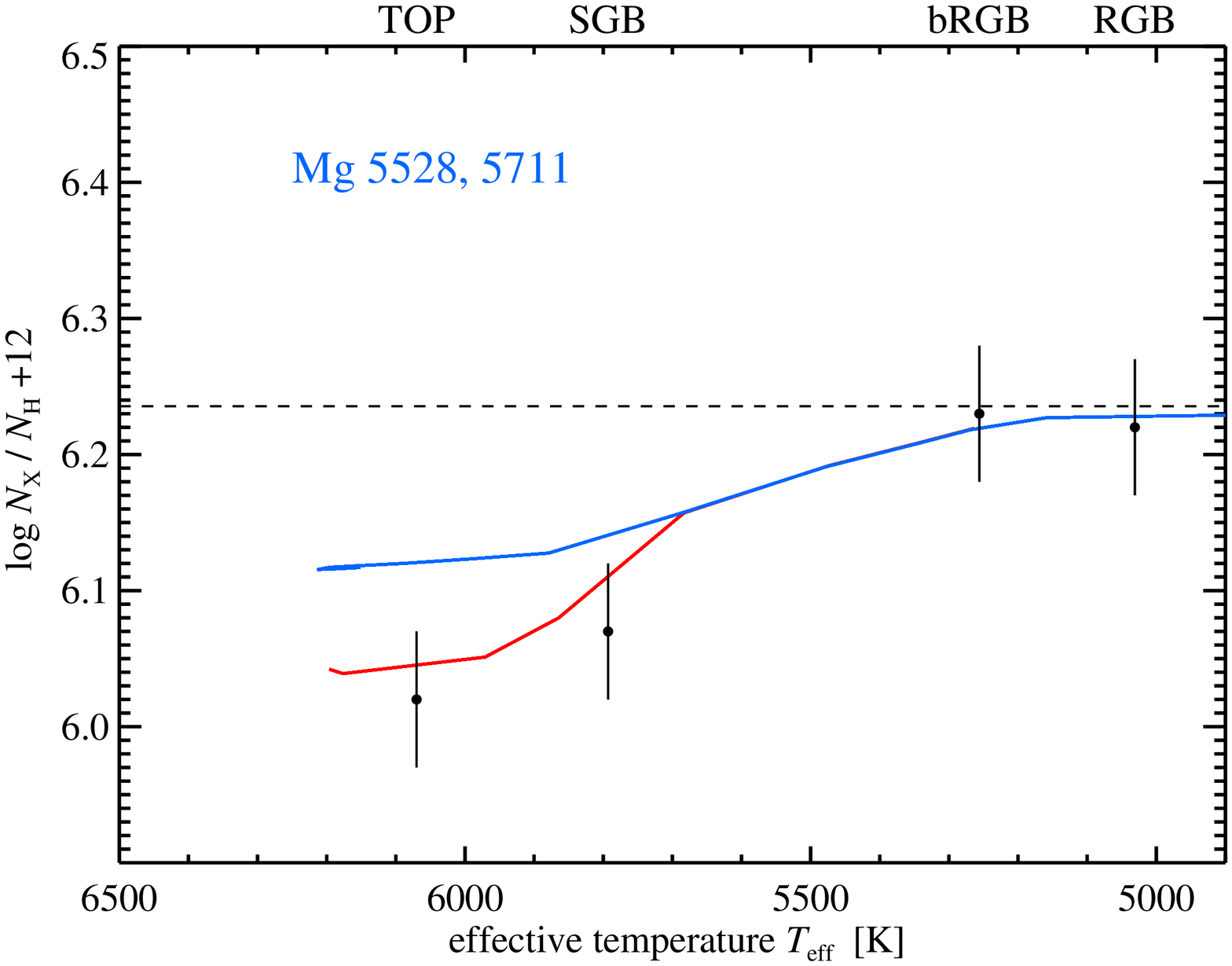}}
\hspace*{-0.06\hsize}
\resizebox{1.09\hsize}{!}{\includegraphics[clip=true, angle=90]{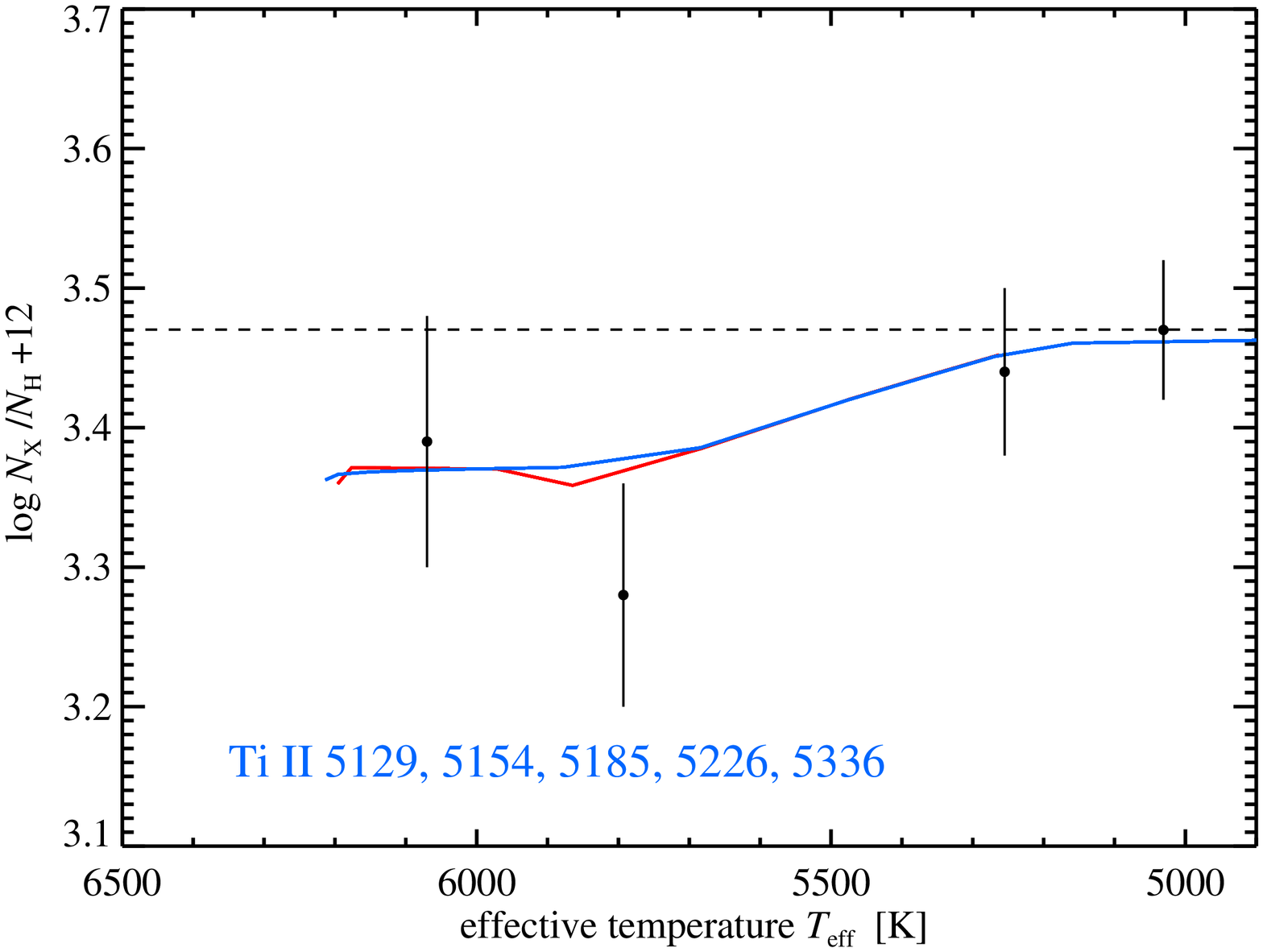}}
\caption{Observed abundance trends for magnesium and titanium in NGC 6752, based on two high-excitation neutral and five ionic lines, respectively. Atomic-diffusion model predictions with NGC 6397-like (T6.0, red) and more efficient (T6.2, blue) extra mixing are overplotted (Richard, priv.\ comm.).
}
\end{figure*}

If atomic diffusion produces sizeable effects in NGC6397, it is not unreasonable to assume that all stars with similar properties are affected. But towards higher metallicities, old stars do not reach as high TOP effective temperatures. And maybe there is an evolution of extra mixing which renders atomic diffusion inefficient.

To investigate this, we have to play the same game as in NGC 6397: carefully analyse a variety of elements in stars between the TOP and the RGB and look for abundance trends. To beat down the error bars, this is best done by looking at ionic lines, as they are predomi\-nantly gravity-sensitive (they have other advantages over neutral lines, like low susceptibility to 3D and NLTE). And relative surface gravities for stars at a common distance can be determined to high precision by combining the observed $V$ magnitude, an assumed distance modulus and estimates of the stellar mass and reddening. The $T_{\rm eff}$ scale has practically no influence on the relative surface-gravity scale.

We are currently finalising the analysis of stars in NGC 6752 at [Fe/H]=$-1.6$. Abundance trends have been checked for iron (from Fe II lines), Ti (from Ti II lines), Sc (from Sc II lines ) and Mg (from high-excitation lines of Mg I which have a low temperature sensitivity). We look at these four elements, as they represent groups of elements which are affected differently by atomic diffusion and extra mixing (due to the element-specific interplay of gravitational settling and radiative acceleration). The resulting trends are shown in Figs.\ 1 \& 2. They seem to indicate that atomic diffusion is operational, but extra mixing is more efficient in this cluster than in NGC 6397. This is very interesting, as the two clusters differ by a mere 0.5\,dex in metallicity.

Mucciarelli et al.\ (2011) looked at M4 ([Fe/H]=$-1.1$) and found absolutely no trend in iron. Still, the lithium abundances require very efficient mixing (T6.25) to be made compatible with WMAP-calibrated BBN. Can simplifying model assumptions (1D, LTE) conspire to hide the small (0.1\,dex) trend in iron expected from a T6.25 Richard model? This is not impossible (I guess I will have to take a look myself). If this result stands the test of time, then we see how atomic diffusion becomes inefficient over only one order of magnitude in metallicity. One wonders immediately what happens at lower metallicities. 

Preliminary results on M 30 ([Fe/H] =$-2.4$) were presented by Karin Lind at this conference. Even lower metallicities cannot be probed with Galactic GCs, this is the realm of very metal-poor field stars where the meltdown of the Spite plateau is observed (Sbordone et al.\ 2010; Aoki, these proceedings; Sbordone, these proceedings). On the one hand, this is a worrying finding for the paradigm of primordial lithium in metal-poor stars. On the other, it may tell us something fundamental about star formation in the early Galaxy.  
  
\section{Should we be worried about Tx.y?}
The extra-mixing efficiency is a free parameter of our present-day modelling capabilities. The pessimistic view is that it removes all predictive power from the atomic-diffusion calculations. The astronomer's point of view is that it is a parameter we can calibrate observationally. Let's face it, many astronomical theories have such free parameters: $\alpha$(mixing-length), micro/mactroturbulence $\xi, \Xi$, neutral-hydrogen collision strength $S_{\rm H}$, you name it. Tx.y is not fundamentally different from these examples. Consequently my answer would be that we should use it, but use it with care; and try to replace it with a physical description as soon as we can. Just like with $\xi, \Xi$ and $S_{\rm H}$, this is in the making (cf.\ Charbonnel, these proceedings).  

\section{Lithium in RGB stars below the bump}
A fresh look at how to best infer the initial lithium abundance of GC stars was taken by Mucciarelli et al.\ (2012). Rather than looking at the Spite plateau itself, they chose to look at RGB stars below the bump. The lithium abundances in these stars can namely be said to form a plateau as well. This has several advantages. One gains two magnitudes in $V$ as these stars are significantly brighter (some of this has to be re-invested into longer exposure times, as the lithium line is weaker than on the Spite plateau). These stars' lithium is furthermore less sensitive to atomic diffusion (0.1\,dex vs.\ 0.3\,dex). But some sensitivity to Tx.y (on the 0.1\,dex level) remains and it seems rather challenging to break this dependence by looking at RGB stars alone. But the ambition is right, one should really aim to understand the whole evolution of lithium.   

\section{Outliers}
Outliers have not received the attention they deserve, some would claim. I would tend to agree. Yet looking at the Spite-plateau stars in NGC 6397 analysed by Lind et al.\ (2009a), all outliers are Na-rich second-generation stars. It seems as if we are not learning something about the distribution in initial angular momentum or other interesting physical parameters in which stars of one and the same population may differ (there are more ``freaks'' among field stars). 

Koch et al.\ (2011) took spectra of a grand total of three TOP stars in NGC 6397 and found one of them to be the most lithium-rich Pop II dwarf to date. Its surface lithium is two orders of magnitude higher than expected. This is an outlier and a half! Who would have guessed that there are unevolved stars with 30 times more lithium than what standard BBN predicts? Modellers with new ideas to the front!

\section{The end of enquiry?}
While some (including me) tend to tone down the remaining cosmological lithium discrepancy, others like to live it up. But does my point of view entail the end of enquiry into lithium? Shall we turn the page and move on? Far from it! As the works covered show, it is merely the focus that has changed. For the time being, we cannot claim to understand what happens with surface lithium as low-mass stars age over billions of years. The proof of concept in favour of atomic diffusion as the cause of a general lowering of surface lithium has been performed, but the parametrized extra mixing makes secure quantitative inferences diffcult at present. While it is, of course, highly desirable to put BBN to the test, the lithium isotopes may just not be the tools to do this right now. But careful studies nontheless allow to trace the surface evolution of lithium and other elements in globular clusters. This will yield additional boundary conditions for models of stellar structure that try to find the right mix of processes giving rise to the extra mixing. A sequence in metallicity (M30, NGC 6397, NGC 6752, M 4) is a step in this direction and first results concerning the evolution of the mixing effciency are promising. 

The last decade has seen the comprehensive application of hydrodynamic (HD) models to abundance work. We can apply NLTE correction with much greater confidence, as input-physics uncertainties related to collisions with neutral hydrogen have been removed, at least for simple atoms like lithium (Barklem, Belyaev \& Asplund 2003). Fabbian et al.\ (2010) explore the effects of magneto-hydrodynamic (MHD) modelling on abundances of Fe I lines in the Sun. On top of the direct Zeeman-splitting effect, one also has an effect stemming from a modified $T-\tau$-relation. It is too early to tell how these effects would affect lithium abundances in metal-poor stars. Assuming a qualitatively similar behaviour as for Fe I lines in the solar photosphere, an increase of the inferred abundance does not seem unlikely. MHD modelling may thus turn into the buzz word for the coming decade.  

\section{Conclusions}
Globular cluster studies of lithium have made very significant contributions to our understanding of surface-lithium evolution in metal-poor stars. I summarize some relevant points:
\begin{itemize}
\item
Atomic diffusion moderated by some form of extra mixing modifies the surface abundances of all Spite-plateau stars, not only but in particular in terms of lithium. The correction is at least +0.2\,dex.
\item
There is no convincing evidence in favour of systematically higher (or lower) lithium abundances of Spite-plateau stars residing in GCs, differences do not seem to exceed 0.05\,dex. Care should, however, be taken to measure lithium in first-generation stars whenever the GC under investigation shows prominent elemental anti-correlations (O-Na, Mg-Al). 
\item
Remaining uncertainties in the absolute effective temperatures of warm halo stars can at most account for a 0.15\,dex shift in log $\varepsilon$(Li). However, the indiect effect of the relative effective-temperature scale on the inferred extra-mixing efficiency and thus on the diffusion correction for lithium can be as large as 
0.4\,dex. Atomic diffusion moderated by efficient extra mixing (T6.25) on a hot absolute effective-temperature scale can fully bridge the gap to the WMAP-calibrated BBN prediction. MHD modelling may alleviate the need for high $T_{\rm eff}$ values.\footnote{I re-iterate my bet from 2010: If you are willing to wager a bottle of wine that the remaining $\leq$ 0.2\,dex between diffusion-corrected stellar lithium and precision-cosmology (WMAP-calibrated BBN) lithium are due to {\em new physics}, contact the author.} 
\item
There is no good reason not to agree on departures from LTE for lithium within a given modelling framework (1D, 3D). For 1D analyses, the Lind et al. (2009b) corrections are recommended.
\item
GC are no direct help in addressing the paradigm-shaking meltdown of the Spite plateau at very low metallicities.
\end{itemize}
What a beautiful mess GC and stellar physics has become during the past decade! 
 
\begin{acknowledgements}
I acknowledge the efforts by and support from my collaborators without whom this work would not have been possible. In particular, I would like to thank Frank Grundahl, Bengt Gustafsson, Lyudmila Mashonkina and Olivier Richard: Your knowledge has been critical in allowing me to see beyond the standard picture. My PhD
student Pieter Gruyters is thanked for proving preliminary results on NGC 6752 prior to publication.
\end{acknowledgements}

\bibliographystyle{aa}

\end{document}